\documentstyle[12pt]{article}

\newcommand {\nn}    {\nonumber}
\newcommand {\vs}[1]  { \vspace*{#1 cm} }

\newcounter{eq}
\newcounter{sc}

\newcommand {\MPL}  {Mod.Phys.Lett.}
\newcommand {\NP}   {Nucl.Phys.}
\newcommand {\PL}   {Phys.Lett.}
\newcommand {\PR}   {Phys.Rev.}
\newcommand {\PRL}   {Phys.Rev.Lett.}

\newcommand {\PTP}  {Prog.Theor.Phys.}

\def\overleftrightarrow#1{\vbox{\ialign{##\crcr
 $\leftrightarrow$\crcr\noalign{\kern-1pt\nointerlineskip}
 $\hfil\displaystyle{#1}\hfil$\crcr}}}

\newlength{\minitwocolumn}
\begin{document}


\begin{flushright}
EDO-EP-18\\
January, 1998\\
\end{flushright}
\vspace{30pt}

\pagestyle{empty}
\baselineskip15pt

\begin{center}
{\large\bf Matrix Theory from Schild Action
 \vskip 1mm
}

\vspace{20mm}

Ichiro Oda
          \footnote{
          E-mail address:\ ioda@edogawa-u.ac.jp
                  }
\\
\vspace{10mm}
          Edogawa University,
          474 Komaki, Nagareyama City, Chiba 270-01, JAPAN \\

\end{center}


\vspace{15mm}
\begin{abstract}
Starting from the Schild action for membrane, we present an alternative
formulation of Matrix Theory. First of all, we construct the
Schild action for general bosonic p-brane which is classically
equivalent to the Nambu-Goto action for p-brane.
Next, based on the constraint obtained from the variational
equation for the auxiliary field in the case of $p = 2$ (membrane), 
we construct a new matrix model which is closely related to the 
matrix model of M-theory as developed by Banks, Fischler, Shenker 
and Susskind (BFSS). 
Our present formulation is a natural extension of the construction 
of type IIB matrix model by Yoneya to the case of M-theory.

\vspace{15mm}

\end{abstract}

\newpage
\pagestyle{plain}
\pagenumbering{arabic}


\rm
\section{Introduction}

Among some dramatic developments in string theory in recent
years maybe the most exciting one has been the discovery
of Matrix Theory \cite{M}. It has long been a mystery to
understand the microscopic degrees of freedom of string 
theory at the short distance regime. For the first time in 
Matrix Theory it was conjectured and further confirmed 
that D-particles \cite{Pol} may be the 
fundamental building block for M-theory \cite{MT}. 
Indeed, it was shown that M-theory is equivalent to
the $N \rightarrow \infty$ limit of the non-relativistic
quantum mechanics of $N$ D-particles in weakly coupled
region of IIA superstring \cite{M}. Thus in Matrix
Theory the starting action for the matrix degrees of 
freedom was derived from IIA superstring.  

On the other hand, there is another matrix model as a
candidate for the non-perturbative formulation of
type IIB superstring, what we call, IIB matrix model
\cite{IKKT}.
The action in this model has the form of the large $N$
matrix model of ten dimensional super Yang-Mills theory 
reduced to a point and was constructed by starting
from the Schild action \cite{Schild} for the 
Green-Schwarz IIB superstring \cite{GS}. 
Furthermore, a similar model to the original IIB 
matrix model was built out of a different method 
where the constraint derived from the variational 
equation with respect to the auxiliary field in 
the Schild action for string has played a critical 
role \cite{Y1}. 
Afterwards, it was also shown 
that this model can be understood from the viewpoint of
breakdown of topological symmetry \cite{Oda1}.

Then one is naturally led to ask whether Matrix 
Theory can be also derived from the Schild action
for membrane by the strategy adopted in the reference
\cite{Y1}. This is the main problem
that I want to study in this paper. We will show that
this is indeed the case. However, in constrast to
the case of IIB matrix model, we have to fix the gauge 
symmetries in the light-cone gauge completely, which 
makes the physically interesting interpretation of 
the constraint as the space-time uncertainty relation 
\cite{Y1, Y2, Amelino} very vague. 

The paper is organized as follows. In section 2 it is shown 
that it is possible to make the Schild action for general
bosonic p-brane that is classically equivalent to the
Nambu-Goto action for p-brane, where special attention is 
paid to the constraints in the Schild action. In section 3,
on a basis of the contraint in the Schild action for
membrane we construct a new matrix model of M-theory 
in the light-cone gauge. It is mentioned that this new
Matrix Theory has $N=2$ supersymmetry and yields
Matrix Theory by Banks et al. \cite{M} as the low energy
effective theory of many distant clusters of D-branes.
The final section is devoted to discussions.

\section{ The Schild action for p-brane }

In this section, we construct the Schild action \cite{Schild} for
general bosonic p-brane that is equivalent to the Nambu-Goto action
for p-brane and then analyse the structure of the constraints in
the Hamiltonian formalism. In this paper, we confine ourselves to
only the equivalence at the classical level since it currently seems
to be very difficult to prove the quantum equivalence among
various formulations of p-brane except in string theory ($p=1$).
Thus, in the following we shall neglect the contribution stemming 
from the functional measures, the ghosts and the normal orderings. 

First of all, let us recall the Schild action \cite{Schild} for
bosonic string ($p = 1$), which is of the form
\begin{eqnarray}
S_n ^{p=1} = -\frac{1}{n} \int d^2 \xi \ e \ 
\left[\frac{1}{e^n} \left\{ -\frac{1}{2 \lambda_1 ^2} 
\left( \sigma^{\mu_1 \mu_2} \right)^2 \right\}^{\frac{n}{2}} 
+ n - 1 \right],
\label{2.1}
\end{eqnarray}
where $e(\xi)$ is a positive definite scalar density defined 
on the string world sheet parametrized by $\xi^0$ and $\xi^1$, 
$\lambda_1 = 2\pi \alpha'$, and $\sigma^{\mu_1 \mu_2}$ is
defined as $\varepsilon^{\alpha_1 \alpha_2} \partial_{\alpha_1}
X^{\mu_1} \partial_{\alpha_2} X^{\mu_2}$. 
Here $X^\mu (\xi)$ $(\mu = 0, 1, \ldots , D-1)$ are space-time 
coordinates and the index $\alpha$ runs over the world sheet
indices 0 and 1. 
Throughout this paper, we assume that
the space-time metric takes the flat Minkowskian form defined as 
$\eta_{\mu\nu} = diag(- + + \cdots +)$.

Then it is quite straightforward to build the Schild action for
general bosonic p-brane by generalizing (\ref{2.1}).
The concrete expression is given by
\begin{eqnarray}
S_n ^p = -\frac{1}{n} \int d^{p+1} \xi \ e \ 
\left[\frac{1}{e^n} \left\{ -\frac{1}{(p+1)! \lambda_p ^2} 
\left( \sigma^{\mu_1 \cdots \mu_{p+1}} \right)^2 \right\}^{\frac{n}{2}} 
+ n - 1 \right],
\label{2.2}
\end{eqnarray}
where $\sigma^{\mu_1 \cdots \mu_{p+1}} = \varepsilon^{\alpha_1 \cdots 
\alpha_{p+1}} \partial_{\alpha_1} X^{\mu_1} \cdots 
\partial_{\alpha_{p+1}} X^{\mu_{p+1}}$ and the world volume index 
$\alpha$ takes the values $0, 1, \cdots, p$. 

In fact, we can demonstrate that (\ref{2.2}) is equivalent to the 
Nambu-Goto action for p-brane as follows. 
Taking the variation with respect to the auxiliary field $e(\xi)$, 
one obtains the constraint
\begin{eqnarray}
e(\xi) = \frac{1}{\lambda_p} \sqrt{-\frac{1}{(p+1)!} 
\left( \sigma^{\mu_1 \cdots \mu_{p+1}} \right)^2}.
\label{2.3}
\end{eqnarray}
Plugging the constraint (\ref{2.3}) into the Schild action (\ref{2.2}),
one obtains
\begin{eqnarray}
S_n ^p &=& -\int d^{p+1} \xi \ e \nn\\
&=& -\frac{1}{\lambda_p} \int d^{p+1} \xi \sqrt{- \det 
\partial_\alpha X^\mu \partial_\beta X_\mu} \nn\\
&=& S_{Nambu-Goto},
\label{2.4}
\end{eqnarray}
where the identity 
\begin{eqnarray}
\det \partial_\alpha X^\mu \partial_\beta X_\mu
= \frac{1}{(p+1)!} \left( \sigma^{\mu_1 \cdots \mu_{p+1}} \right)^2
\label{2.5}
\end{eqnarray}
was used. Hence the Schild action (\ref{2.2}) becomes at least 
classically equivalent to the famous form of the Nambu-Goto action 
$S_{NG}$.

In order to understand the constraint (\ref{2.3}) more closely,
it is useful to make use of the Hamiltonian formalism. 
The canonical conjugate momenta to the $X^\mu$ are given by
\begin{eqnarray}
P_\mu &=& \frac{1}{e^{n-1}} \frac{1}{p! \lambda_p ^2} 
\left\{ -\frac{1}{(p+1)! \lambda_p ^2} 
\left( \sigma^{\mu_1 \cdots \mu_{p+1}} \right)^2 
\right\}^{\frac{n}{2}-1} \nn\\
& & {} \times \sigma_{\mu \mu_1 \cdots \mu_p}
\varepsilon^{i_1 \cdots i_p} 
\partial_{i_1} X^{\mu_1} \cdots 
\partial_{i_p} X^{\mu_p},
\label{2.6}
\end{eqnarray}
where the index $i$ takes the values from 1 to p. From (\ref{2.6}),
it is easy to see that the momenta satisfy the primary constraints
\begin{eqnarray}
P_\mu \partial_i X^\mu &=& 0, \\
P^2 + \frac{1}{\lambda_p ^2} \det \partial_i X^\mu 
\partial_j X_\mu &=& 0,
\label{2.7}
\end{eqnarray}
where the lapse (Hamiltonian) constraint (\ref{2.7}) is a 
consequence of the constraint (\ref{2.3}) while the shift (momentum) 
constraints (7) come from the definition (\ref{2.6}) trivially.
In this sense, the constraint (\ref{2.3}) encodes all the dynamical
informations of the Schild action for p-brane. 

At this stage, it is valuable to point out that in the case of 
string theory the constraint (\ref{2.3}) expresses the space-time 
uncertainty principle of string \cite{Y2} when the Poisson bracket 
is replaced by a commutator in the large $N$ matrix model, and is 
utilized as the first principle for constructing a type 
IIB supersymmetric matrix model \cite{Y1}. Then it is quite natural 
to ask whether one can also construct a new matrix theory of M-theory
if one starts with the constraint (\ref{2.3}) in the case of membrane
$p=2$. In the next section, we shall show that this conjecture is 
indeed true, but to this aim we have to fix the gauge symmetries
in terms of the light-cone gauge completely.

\section{ Matrix Theory from Schild action for membrane }

The Schild action for string \cite{Schild} has provided us with 
a useful starting point for constructing the matrix models 
of the type IIB superstring \cite{IKKT, Y1}. Since it was 
shown in the previous section that we can also construct the Schild 
action for general p-brane, a natural question arises as to whether
it is possible to implement an analogous procedure in the 
M-theory, which amounts to making Matrix Theory by starting with 
the Schild action for membrane.
This is exactly the problem that I want to address in 
this article. 

Hence let us start with the Schild action for bosonic membrane which
is obtained by putting $p=2$ in (\ref{2.2}), in eleven
space-time dimensions $D=11$. 
Then the action (\ref{2.2}) becomes
\begin{eqnarray}
S_n ^{p=2} = -\frac{1}{n} \int d^3 \xi \ e \ 
\left[\frac{1}{e^n} \left\{ -\frac{1}{3! \lambda_2 ^2} 
\left( \sigma^{\mu \nu \rho} \right)^2 \right\}^{\frac{n}{2}} 
+ n - 1 \right],
\label{3.1}
\end{eqnarray}
and the constraint (\ref{2.3}) is given by
\begin{eqnarray}
e(\xi) &=& \frac{1}{\lambda_2} \sqrt{-\frac{1}{3!} 
\left( \sigma^{\mu \nu \rho} \right)^2} \nn\\
&=& \frac{1}{\lambda_2} \sqrt{ 
- \det \partial_\alpha X^\mu \partial_\beta X_\mu},
\label{3.2}
\end{eqnarray}
where the constant $\lambda_2 = l_{11}^3$ with the eleven
dimensional Planck length $l_{11}$ is the inverse of
the membrane tension, which will set to be a unity
from now on for simplicity. Here for later convenience,
it is useful to cast the above constraint (\ref{3.2})
into an alternative form
\begin{eqnarray}
e(\xi) = \sqrt{ g \ \Delta },
\label{3.3}
\end{eqnarray}
where we introduced the notations \cite{de Witt}
\begin{eqnarray}
g = \displaystyle{ \det_{i,j=1,2} g_{ij}} 
= \displaystyle{ \det_{i,j=1,2} \partial_i X^\mu 
\partial_j X_\mu}, \nn\\
\Delta = - g_{00} + u^i u_i 
= - \left( (\partial_0 - u^i \partial_i) X^\mu \right)^2, \nn\\
u_i = g_{0i}, \ u^i = g^{ij} u_j.
\label{3.4}
\end{eqnarray}

At this point, let us give up the Lorentz covariance
and take the light-cone gauge. This is due to the fact that 
it seems to be difficult to apply the Goldstone-Hoppe map
between representation theory of the algebra of the area-preserving
diffeomorphisms and the large $N$ matrix theory \cite{Hoppe},
whose map will be crucially utilized later in constructing the matrix
theory.
The light-cone gauge in the membrane theory is chosen
to be \cite{de Witt}
\begin{eqnarray}
\partial_\alpha X^+ = \delta_{\alpha 0}, \ 
u^i = 0, 
\label{3.5}
\end{eqnarray}
where the light-cone coordinates are defined as 
$X^{\pm} = \frac{1}{\sqrt 2} ( X^{10} \pm X^0 )
= X_{\mp}$. Note that in the light-cone gauge, $g$ and 
$\Delta$ in (\ref{3.4}) become
\begin{eqnarray}
g = \displaystyle{ \det_{i,j=1,2} \partial_i X^a 
\partial_j X^a}
=\frac{1}{2} \{ X^a, X^b \}^2, \nn\\
\Delta = - g_{00} = - ( \partial_0 X^\mu )^2, 
\label{3.6}
\end{eqnarray}
where $a, b = 1, \cdots, 9$ stand for the indices of the 
transverse coordinates, and the curly bracket denotes the 
Poisson bracket defined by $\{ A, B \} = \varepsilon^{ij}
\partial_i A \partial_j B$. 

It is well known that the membrane theory has 
the area-preserving diffeomorphisms as residual 
symmetries of the world volume diffeomorphisms 
in the light-cone gauge \cite{de Witt}. 
Since our idea in this paper is to fix the gauge
symmetries completely, the area-preserving 
diffeomorphisms are also used to pick a gauge 
\begin{eqnarray}
e(\xi) = \sqrt{ \Delta \left( \partial_0 X^a 
\right)^2}. 
\label{3.7}
\end{eqnarray}

Under these gauge conditions, the constraint (\ref{3.3}) 
in which we are mainly interested is simply reduced to
\begin{eqnarray}
\frac{1}{2} \left( \partial_0 X^a \right)^2
- \frac{1}{4} \{ X^a, X^b \}^2 = 0. 
\label{3.8}
\end{eqnarray}
Our basic idea for the construction of the matrix model of M-theory,
that is, Matrix Theory
is to take the condition (\ref{3.8}) as the fundamental condition.
This strategy was adopted by Yoneya \cite{Y1} to build the matrix
model of type IIB superstring theory, IIB matrix model,
since the corresponding constraint in string theory realizes his 
space-time uncertainty relation \cite{Y1, Y2}. Here it is worthwhile
to comment on the difference between string and membrane.
In string theory, 
it is transparent that the constraint (\ref{2.3}) with $p=1$ 
describes the space-time uncertainty relation $\Delta T \Delta X
\ge l_s$ in the matrix theory \cite{Y1}, while in 
membrane theory it is at present unclear whether the corresponding
constraint (\ref{3.2}) (or (\ref{3.3})) or its gauge fixed version
(\ref{3.8}) expresses the space-time uncertainty 
relation or not. This problem is closely related to
our gauge choice where the light-cone gauge kills the dynamical
degrees of freedom associated with $X^+$ or $X^0$ by choosing the
gauge $\partial_\alpha X^+ = \delta_{\alpha 0}$. The space-time
uncertainty relation advocated in \cite{Y1, Y2} is the statement
about the relation between the time component $X^0$ or $X^+$
and one space component $X^a$ so that it is obvious that we cannot
understand the space-time uncertainty relation within the present
formulation.  
Perhaps it might be an important step
towards the covariant formulation of Matrix Theory that 
we try to understand the relationship between the constraint
(\ref{3.2}) and the space-time uncertainty relation
of membrane without taking the light-cone gauge.  
In this note, without worrying such a difficult problem,
let us follow the strategy mentioned above and ask if we can
make Matrix Theory on a basis of the gauge-fixed constraint
(\ref{3.8}). 

Before doing so, let us use the Goldstone-Hoppe prescription 
given by \cite{Hoppe}
\begin{eqnarray}
\{ X^a, X^b \} &\rightarrow&  \frac{1}{i} [ X^a, X^b ], \nn\\
\int d\xi^1 d\xi^2 &\rightarrow& Tr.
\label{3.9}
\end{eqnarray}
Then the constraint (\ref{3.8}) becomes
\begin{eqnarray}
\frac{1}{2} \left( \partial_0 X^a \right)^2
+ \frac{1}{4} [ X^a, X^b ]^2 = 0. 
\label{3.10}
\end{eqnarray}
Provided that we require that a weaker form of this constraint
\begin{eqnarray}
\int d\xi^0 Tr \left( \frac{1}{2} \left( \partial_0 X^a \right)^2
+ \frac{1}{4} [ X^a, X^b ]^2 \right) = 0 
\label{3.11}
\end{eqnarray}
is the fundamental condition for the construction of Matrix Model,
the partition function can be defined as 
\begin{eqnarray}
Z = \int {\it D} X^a {\it J}[X] \ \delta 
\left( \int d\xi^0 Tr \left( \frac{1}{2} 
\left( \partial_0 X^a \right)^2
+ \frac{1}{4} [ X^a, X^b ]^2 \right) \right), 
\label{3.12}
\end{eqnarray}
where ${\it J}[X]$ is a certain measure factor which will be 
determined by following the similar consideration as done in the 
reference \cite{Y1} whose result is given by
\begin{eqnarray}
{\it J}[X] = \int {\it D} \theta \exp \int d\xi^0 Tr
\left( - \theta^T \partial_0 \theta - \theta^T 
\gamma_a [ \theta, X^a ] \right). 
\label{3.13}
\end{eqnarray}
After all, the partition function has the form
\begin{eqnarray}
Z &=& \int {\it D} c {\it D} X^a {\it D} \theta 
\exp \Bigl[  c \int d\xi^0 Tr \left( \frac{1}{2} 
\left( D_0 X^a \right)^2
+ \frac{1}{4} [ X^a, X^b ]^2 \right)  \nn\\
& & {} - \int d\xi^0 Tr \left( \theta^T D_0 \theta + \theta^T 
\gamma_a [ \theta, X^a ]  \right)
\Bigr], 
\label{3.14}
\end{eqnarray}
where the covariant derivative $D_0 = \partial_0 + i A_0
$ was introduced to guarantee the closure under the 
$N=2$ supersymmetry
\begin{eqnarray}
\delta X^a &=& -2 \epsilon^T \gamma^a \theta, \\
\delta \theta &=& \frac{c}{2} \left( D_0 X^a \gamma_a + 
\gamma_- + \frac{1}{2} [ X^a, X^b ] \gamma_{ab} \right) 
\epsilon + \epsilon^\prime, \\
\delta A_0 &=& -2 \epsilon^T \theta, \\
\delta c &=& 0,
\label{3.15}
\end{eqnarray}
where $\epsilon$ and $\epsilon^\prime$ are two independent 
spinorial parameters. In this way, we have succeeded in 
deriving a Matrix Theory with $N=2$ supersymmetry by
starting with the Schild action for membrane. Of course,
following the line of similar arguments to
in the reference \cite{Y1}, we can show that the original 
Matrix Theory \cite{M} can be interpreted as the low energy
effective theory of many distant clusters of D-branes from
this Matrix Theory.

\section{ Discussions }
In this paper, we have pursued the possibility of formulating
Matrix Theory in terms of the Schild action for membrane.
We have seen that in order to carry out this idea we have
to fix the gauge symmetries in the light-cone gauge and
furthermore fix the residual area-preserving diffeomorphisms
completely. The constraint in the Schild action for membrane,
which we have adopted as the fundamental condition, 
encodes all the dymanical informations of the system, but
it is not clear that this constraint has something to
do with the space-time uncertainty principle adovacated
by Yoneya \cite{Y1, Y2} because of the light-cone gauge
choice. To fully understand the details of it, we have to
wait for completion of the covariant formulation of Matrix 
Theory.

However, we should remark that no Lorentz covariant 
formulation can be more than a step on the road to
the non-perturbative formulation of M-theory. This is
because whatever it is, M-theory cannot have its most
fundamental formulation on the fixed, flat Minkowski
background mainifold. In other words, M-theory should
be independent of the background metric in the 
most fundamental form, and the background metric must
emerge from its fundamental theory through some
unknown dynamical mechanism. Recently, such matrix models
which do not depend on the background metric have
been pushed forward \cite{Smolin, Oda2}.

\vs 1
\begin{flushleft}
{\bf Acknowledgement}
\end{flushleft}
The author thanks A.Sugamoto for valuable discussions
and continuous encouragement. 
This work was supported in part by Grant-Aid 
for Scientific Research from Ministry of Education, Science and
Culture No.09740212.

\vs 1


\begin{thebibliography}{99}
\bibitem{M}
T.Banks, W.Fischler, S.H.Shenker and L.Susskind, {\PR{\bf D55}
(1997) 5112.}
\bibitem{Pol}
J.Polchinski, {\PRL{\bf 74} (1995) 4724.}
\bibitem{MT}
C.M.Hull and P.K.Townsend, {\NP{\bf B438} (1995) 109; 
E.Witten, \NP{\bf 443} (1995) 85.}
\bibitem{IKKT}
N.Ishibashi, H.Kawai, Y.Kitazawa and A.Tsuchiya, 
{\NP{\bf B498} (1997) 467.}
\bibitem{Schild}
A.Schild, {\PR{\bf D16} (1977) 1722.}
\bibitem{GS}
M.B.Green and J.H.Schwarz, {\PL{\bf B136} (1984) 367.}
\bibitem{Y1}
T.Yoneya, {\PTP{\bf 97} (1997) 949.}
\bibitem{Oda1} 
I.Oda, {hep-th/9709005}, {\MPL{\bf A}} (in press); 
{hep-th/9710030}, {\NP{\bf B}} (in press).
\bibitem{Y2}
T.Yoneya, {\MPL{\bf A4} (1989) 1587}; M.Li and T.Yoneya, {\PRL{\bf 78}
(1997) 1219.}
\bibitem{Amelino} 
G. Amelino-Camelia, J.Lukierski and A. Nowicki, 
{hep-th/9706031} and references therein.
\bibitem{de Witt}
B.de Witt, J.Hoppe and N.Nicolai, {\NP{\bf B305 [FS23]} (1988) 
545.}
\bibitem{Hoppe}
J.Hoppe, {MIT Ph.D Thesis, 1982, published in Soryuushiron
Kenkyu (Kyoto) {\bf 80} (1989) 145.}
\bibitem{Smolin}
L.Smolin, {hep-th/9710191.}
\bibitem{Oda2} 
I.Oda, {hep-th/9801051.}

\end{thebibliography}
\end{document}